\documentclass[iop,apjl]{emulateapj}
\usepackage{natbib}
\usepackage{amsfonts}
\usepackage{amssymb}

\shorttitle{parity asymmetry of WMAP power spectrum}
\shortauthors{Jaiseung Kim and et al.}

\begin{document}
\title{Anomalous parity asymmetry of the WMAP power spectrum data at low multipoles}
\author{Jaiseung Kim and Pavel Naselsky}
\affil{Niels Bohr Institute \& Discovery Center, Blegdamsvej 17, DK-2100 Copenhagen, Denmark}
\email{jkim@nbi.dk}
\begin{abstract}
We have investigated non-Gaussianity of our early Universe by comparing the parity asymmetry of the WMAP power spectrum with simulations.
We find that odd-parity preference of the WMAP data ($2\le l\le 18$) is anomalous at 4-in-1000 level. We find it likely that low quadrupole power is part of this parity asymmetry rather than an isolated anomaly. Futher investigation is required to find out whether the origin of this anomaly is cosmological or systematic effect. 
The data from Planck surveyor, which has systematics distinct from the WMAP, will help us to resolve the origin of the anomalous odd-parity preference.
\end{abstract}


\section{Introduction}
For the past years, there have been great successes in measurement of Cosmic Microwave Background (CMB) anisotropy by ground and satellite observations \citep{WMAP5:basic_result,WMAP5:powerspectra,WMAP5:parameter,ACBAR,ACBAR2008,QUaD1,QUaD2,QUaD:instrument,QUaD_improved}.
Recently, Planck surveyor has been successfully launched, and is measuring CMB temperature and polarization anisotropy with very fine angular resolution. Using CMB data, we may test cosmological hypotheses and impose significant constraints on cosmological models \citep{Modern_Cosmology,Inflation,Foundations_Cosmology}.
For the past years, WMAP data have gone through scrutiny, and various anomalies have been reported \citep{cold_spot1,cold_spot2,cold_spot_wmap3,cold_spot_origin,Tegmark:Alignment,Multipole_Vector1,Multipole_Vector2,Multipole_Vector3,Multipole_Vector4,Axis_Evil,Axis_Evil2,Axis_Evil3,Park_Genus,Chiang_NG,PMF1_WMAP1,Hemispherical_asymmetry,power_asymmetry_subdegree,power_asymmetry_wmap5,alfven_Kim}. 
In direct relevance to this letter, Land and et al. have noted odd point-parity preference in WMAP data, but found its statistical significance was not high, given their estimator \citep{Universe_odd}. In this letter, we revisit the point-parity of the WMAP data with a slightly different estimator, and report 
the odd-parity preference of the WMAP power spectrum data at 99.6\% level.
\section{Analysis of the WMAP data}
For a whole-sky CMB analysis, temperature anisotropy $T(\theta,\phi)$ is conveniently decomposed in terms of spherical harmonics $Y_{lm}(\theta,\phi)$ :
\begin{eqnarray*}
T(\hat{\mathbf n})=\sum_{lm} a_{lm}\,Y_{lm}(\hat{\mathbf n}),
\end{eqnarray*}
where $a_{lm}$ is a decomposition coefficient, and $\hat{\mathbf n}$ is a sky direction.
For a Gaussian seed fluctuation model, decomposition coefficients satisfy the following statistical properties:
\begin{eqnarray*} 
\langle a_{lm} \rangle &=& 0 \\
\langle a^*_{lm} a_{l'm'} \rangle &=& C_l\,\delta_{ll'}\delta_{mm'},
\end{eqnarray*}
where $\langle\ldots\rangle$ denotes the average over the ensemble of universes.
\begin{figure}[htb!]
\centering\includegraphics[scale=.5]{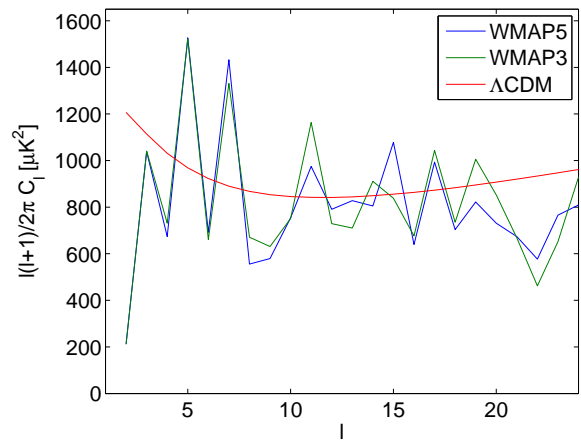}
\caption{CMB power spectrum: $\Lambda$CDM model (red), WMAP 5 year data (blue), WMAP 3 year data (green)}
\label{Cl_cut}
\end{figure}
Given a standard cosmological model, we expect Sach-Wolf plateau for CMB power spectrum on low multipoles \citep{Modern_Cosmology}:
\begin{eqnarray}
l(l+1) C_l\sim \mathrm{const}.\label{plateau}
\end{eqnarray}
In Fig. \ref{Cl_cut}, we show the WMAP 5 year, 3 year data and the theoretical power spectrum of the WMAP concordance model \citep{WMAP3:temperature,WMAP5:powerspectra,WMAP5:Cosmology}. In comparison with WMAP 3 year data, WMAP 5 year data is expected to have more accurate calibration and less foreground contamination \citep{WMAP5:basic_result,WMAP5:powerspectra,WMAP5:beam}. 

Spherical harmonics behave under parity inversion as follows  \citep{Arfken}: 
$Y_{lm}(\hat{\mathbf n})=(-1)^l\,Y_{lm}(-\hat{\mathbf n})$.
Therefore, power asymmetry between even and odd multipoles may be thought as power asymmetry between even and odd parity map, because a map consisting of even(odd) multipoles possesses even(odd) parity. Hereafter, we will denote it as `parity asymmetry'.
\begin{figure}[htb]
\centering\includegraphics[scale=.5]{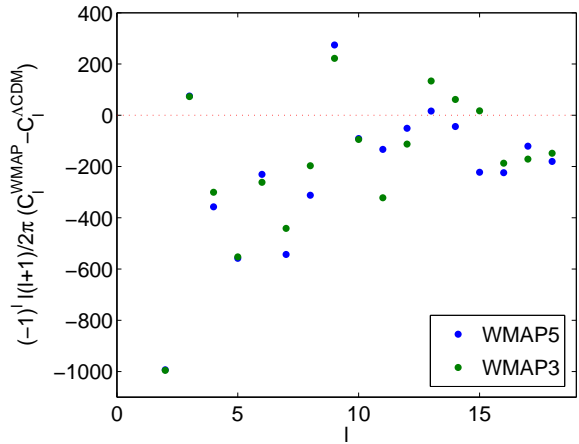}
\caption{$(-1)^l\times$ difference between WMAP power spectrum data and $\Lambda$CDM model}
\label{delta}
\end{figure}
In Fig. \ref{delta}, we show $(-1)^l l(l+1)/2\pi\:(C^{\mathrm{WMAP}}_l-C^{\Lambda\mathrm{CDM}}_l)$ at low multipoles.
As shown in Fig. \ref{delta}, most of them posses negative values, which indicates there exist power deficit (excess) in comparison to the $\Lambda$CDM model at most of even (odd) multipoles. In the case of WMAP5 data, there is only 3 points of positive values among 18 data points.
A order-of-magnitude estimation shows that such events require the odd of $18!/(3!\,15!\:2^{18})\approx 0.003$. 
However, power spectrum is estimated from cut-sky data to avoid diffuse Galactic foreground contamination.
Therefore, statistical fluctuation in estimated $C_l$ is correlated among multipoles.
In order to investigate odd of the parity asymmetry rigorously, we have produced $10^4$ simulated CMB maps (HEALPix Nside=8) of Gaussian $\Lambda$CDM model.
We have degraded the WMAP processing mask (Nside=16) to Nside=8, and set pixels to zero, if any of their daughter pixels is zero. 
After applying the mask, we have estimated power spectrum from cut-sky maps by a pixel-based Maximum-Likelihood method.
Instrument noise is neglected in simulation, since noise is subdominant on multipoles of interest (e.g. S/N $\sim100$ for $C_l$ at $l=30$) \citep{WMAP5:powerspectra}.
Bearing Eq. \ref{plateau} in mind, we consider the following quantities:
\begin{eqnarray} 
P^{+} &=& \sum (l+1-2\left\lfloor \frac{l+1}{2}\right\rfloor)\: l(l+1)/2\pi \: C_l\\
P^{-} &=& \sum (l-2\left\lfloor \frac{l}{2}\right\rfloor)\: l(l+1)/2\pi \: C_l
\end{eqnarray}
where $\lfloor\,\cdots\rfloor$ denotes the greatest integer smaller than or equal to the argument.
Using the WMAP power spectrum data and simulations respectively, we have computed the ratio $P^+/P^-$ for various multipole ranges $2\le l \le l_{\mathrm{max}}$, where $l_{\mathrm{max}}$ is between 3 and 23. By comparing $P^+/P^-$ of the WMAP data with simulation,
we have estimated $p$-value, where $p$-value denotes fractions of simulations as low as $P^+/P^-$ of the WMAP data.
\begin{figure}[htb]
\centering\includegraphics[scale=.3]{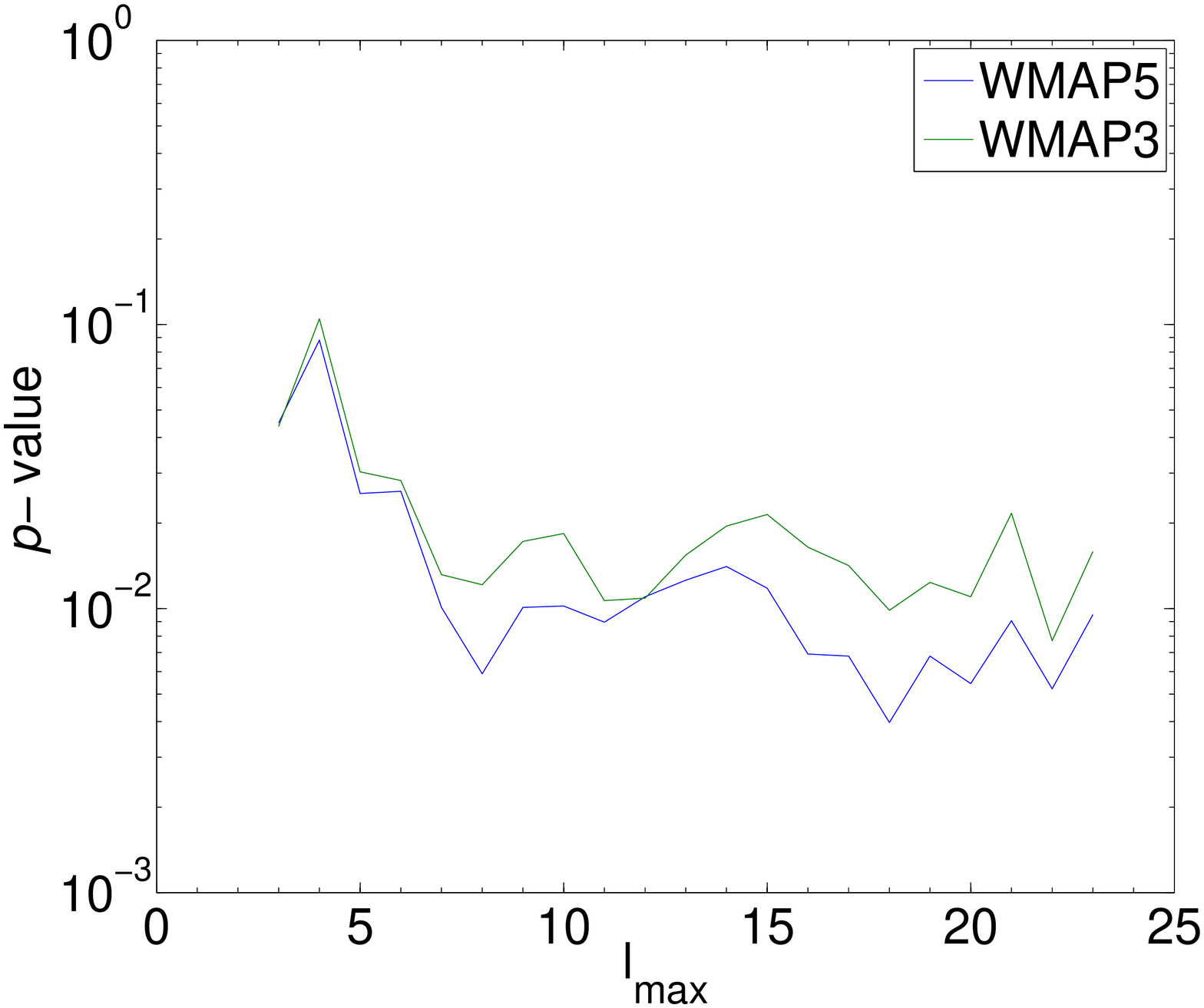}
\caption{Probability of getting $P^+/P^-$ as low as WMAP data for multipole range $2\le l\le{l_\mathrm{max}}$.}
\label{pvalue}
\end{figure}
In Fig. \ref{pvalue}, we show $p$-value of WMAP5 and WMAP3 respectively for various $l_{\mathrm{max}}$.
Fig. \ref{pvalue} shows lowest $p$-value for $l_{\mathrm{max}}=18$, where $p$-values are $0.004$ and $0.0099$ for WMAP5 and WMAP3 respectively. 
In other words, there exists anomalous odd-parity preference at multipoles ($2\le l\le 18$). 
As shown in Fig. \ref{pvalue}, WMAP5 possesses more anomalous odd-parity preference than WMAP3, while WMAP5 data have more accurate calibration and less foreground contamination \citep{WMAP5:basic_result,WMAP5:beam,WMAP5:powerspectra}. Therefore, we find it unlikely that calibration or foregrounds are the source of the anomaly. 
It should be also noted that the anomaly is associated with the WMAP power spectrum data, in which most efforts have been exerted to minimize systematics.

\begin{figure}[htb]
\centering\includegraphics[scale=.5]{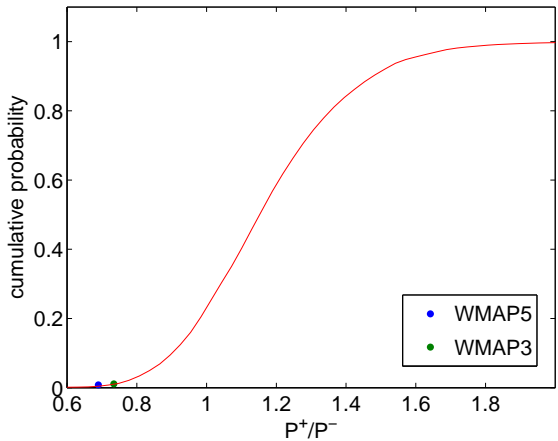}
\caption{Parity asymmetry at multipoles ($2\le l\le 18$): cumulative distribution of $P^+/P^-$ for $10^4$ simulated maps (red), $P^+/P^-$ of WMAP5 (blue) and WMAP3 (green)}
\label{hist}
\end{figure}

It has been known that CMB quadrupole power of WMAP data is unusually low, compared with the theoretical value \citep{Tegmark:Alignment}.
Therefore, one may attribute the anomalous parity asymmetry of the WMAP data to low quadrupole power.
As shown in Fig. \ref{pvalue}, the parity asymmetry persists over extended range of multipoles, and the parity asymmetry on multipoles ($2\le l\le 18$) is most anomalous.
Therefore, we may not simply attribute the parity asymmetry to low quadrupole power.
For multipole range ($2\le l\le 18$), we find $P^+/P^-\approx 1.1$ is most likely, while $P^+/P^-$ of WMAP5 and WMAP3 are 0.69 and 0.734 respectively.
In Fig. \ref{hist}, we show $P^+/P^-$ values of WMAP data and cumulative distribution of $P^+/P^-$ for $10^4$ simulated maps.

We have also estimated $p$-value, using whole-sky simulation (i.e. no mask), and obtained 0.0024.
The difference from the cut-sky result is attributed to the increased statistical fluctuation in cut-sky $C_l$ estimation.
By using whole-sky simulations, we have also investigated $p$-value for $l_{\mathrm{max}}\gg 23$, but have not found the statistical significance as high as $l_{\mathrm{max}}=18$.

\section{Discussion}
In the previous study \citep{Universe_odd}, the parity asymmetry under point reflection as well as mirror reflection was noted, but point-parity was not given enough attention, since they found the statistical significance was not high.
Investigating the WMAP power spectrum with a slightly different estimator, we found the odd-parity preference of the WMAP data ($2\le l\le 18$) at 99.6\% level (mask) and at 99.76\% level (no mask). 
Higher parity asymmetry in WMAP5 data indicates that WMAP systematics is unlikely to be the source for the parity asymmetry.
However, we do not completely rule out non-cosmological origins, and defer a rigorous investigation on cosmological or non-cosmological origin to a separate publication.

One may attribute low $P^+/P^-$ of the WMAP data simply to low quadrupole power.
However, as shown in Fig. \ref{pvalue}, the anomalous parity asymmetry (i.e. low $p$-value) persists over extended range of multipoles.
Therefore, we find it rather likely that low quadrupole power is part of this parity asymmetry anomaly.
It was also shown that hemispherical power asymmetry is much more anomalous at multipoles ($2\le l\le 19$) than multipoles ($20\le l\le 40$) \citep{Hemispherical_asymmetry}.
Given all these circumstantial evidences, we find it likely that there exists an underlying common origin for the anomalies (e.g. hemispherical power asymmetry, low quadrupole power and parity asymmetry), whether it may be cosmological or WMAP systematics.

We are grateful to an anonymous referee for helpful comments, which leads to improvements of this letter.
We are grateful to Changbom Park, Gary Hinshaw and Glenn Starkman for useful discussion.
We acknowledge the use of the Legacy Archive for Microwave Background Data Analysis (LAMBDA). 
Our data analysis made the use of HEALPix \citep{HEALPix:Primer,HEALPix:framework}.
This work is supported in part by Danmarks Grundforskningsfond, which allowed the establishment of the Danish Discovery Center.
This work is supported by FNU grant 272-06-0417, 272-07-0528 and 21-04-0355. 
This work is supported in part by Danmarks Grundforskningsfond, which allowed the establishment of the Danish Discovery Center.

\end{document}